\documentclass{article}

\makeatletter
[1996/3/25 v1.0; Macros for Young-Tableaux]
\newif\ify@autoscale \y@autoscaletrue \def\Yautoscale#1{\ifnum #1=0
  \y@autoscalefalse\else\y@autoscaletrue\fi}
\newdimen\y@b@xdim
\newdimen\y@boxdim \y@boxdim=13pt
\def\Yboxdim#1{\y@autoscalefalse\y@boxdim=#1}
\newdimen\y@linethick    \y@linethick=.3pt
\def\Ylinethick#1{\y@linethick=#1}
\newskip\y@interspace \y@interspace=0ex plus 0.3ex
\def\Yinterspace#1{\y@interspace=#1}
\newif\ify@vcenter   \y@vcenterfalse
\def\Yvcentermath#1{\ifnum #1=0 \y@vcenterfalse\else\y@vcentertrue\fi}
\newif\ify@stdtext   \y@stdtextfalse
\def\Ystdtext#1{\ifnum #1=0 \y@stdtextfalse\else\y@stdtexttrue\fi}
\relax
\def\y@vr{\vrule height0.8\y@b@xdim width\y@linethick depth 0.2\y@b@xdim}
\def\y@emptybox{\y@vr\hbox to \y@b@xdim{\hfil}}
\def\y@abcbox#1{\y@vr\hbox to \y@b@xdim{\hfil#1\hfil}}
\def\y@mathabcbox#1{\y@vr\hbox to \y@b@xdim{\hfil$#1$\hfil}}
\def\y@setdim{%
  \ify@autoscale%
    \ifvoid1\else\typeout{Package youngtab: box1 not free! Expect an
      error!}\fi%
    \setbox1=\hbox{A}\y@b@xdim=1.6\ht1 \setbox1=\hbox{}\box1%
  \else\y@b@xdim=\y@boxdim \advance\y@b@xdim by -2\y@linethick
  \fi}
\newcount\y@counter
\newif\ify@islastarg
\def\y@lastargtest#1,#2 {\ifcat\space #2 \y@islastargtrue
  \else\y@islastargfalse\fi}
\def\y@emptyboxes#1{\y@counter=#1\loop\ifnum\y@counter>0
  \advance\y@counter by -1 \y@emptybox\repeat}
\def\y@nelineemptyboxes#1{%
  \vbox{%
    \hrule height\y@linethick%
    \hbox{\y@emptyboxes{#1}\y@vr}
    \hrule height\y@linethick}\vspace{-\y@linethick}}
\def\yng(#1){%
  \y@setdim%
  \hspace{\y@interspace}%
  \ifmmode\ify@vcenter\vcenter\fi\fi{%
  \y@lastargtest#1,
  \vbox{\offinterlineskip
    \ify@islastarg
      \y@nelineemptyboxes{#1}
    \else
      \y@ungempty(#1)
    \fi}}\hspace{\y@interspace}}
\def\y@ungempty(#1,#2){%
  \y@nelineemptyboxes{#1}
  \y@lastargtest#2,
  \ify@islastarg
    \y@nelineemptyboxes{#2}
  \else
    \y@ungempty(#2)
  \fi}
\def\y@nelettertest#1#2. {\ifcat\space #2 \y@islastargtrue
  \else\y@islastargfalse\fi}
\def\y@abcboxes#1#2.{%
  \ify@stdtext\y@abcbox#1\else\y@mathabcbox#1\fi%
  \y@nelettertest #2.
  \ify@islastarg\unskip%
  \ify@stdtext\y@abcbox{#2}\else\y@mathabcbox{#2}\fi%
  \else\y@abcboxes#2.\fi}
\def\y@nelineabcboxes#1{%
  \y@nelettertest #1.
  \ify@islastarg
    \vbox{
      \hrule height\y@linethick%
      \hbox{\ify@stdtext\y@abcbox#1\else\y@mathabcbox#1\fi\y@vr}
      \hrule height\y@linethick}\vspace{-\y@linethick}
  \else \vbox{
      \hrule height\y@linethick%
      \hbox{\y@abcboxes #1.\y@vr}%
      \hrule height\y@linethick}\vspace{-\y@linethick}
  \fi}
\def\young(#1){%
  \y@setdim%
  \hspace{\y@interspace}%
  \y@lastargtest#1,
  \ifmmode\ify@vcenter\vcenter\fi\fi{%
  \vbox{\offinterlineskip
    \ify@islastarg\y@nelineabcboxes{#1}%
    \else\y@ungabc(#1)%
    \fi}}\hspace{\y@interspace}}
\def\y@ungabc(#1,#2){%
  \y@nelineabcboxes{#1}%
  \y@lastargtest#2,
  \ify@islastarg\y@nelineabcboxes{#2}%
  \else\y@ungabc(#2)%
  \fi}
\makeatother

\newcommand{\be}{\begin{equation}}
\newcommand{\ee}{\end{equation}}
\newcommand{\ba}{\begin{eqnarray}}
\newcommand{\ea}{\end{eqnarray}}
\newcommand{\bas}{\begin{eqnarray*}}
\newcommand{\eas}{\end{eqnarray*}}

\newcommand{\nn}{\nonumber}

\newcommand{\gam}{\gamma}
\newcommand{\Gam}{{\mit \Gamma}}

\newcommand{\lam}{\lambda}

\newcommand{\cA}{{\cal A}}
\newcommand{\cB}{{\cal B}}
\newcommand{\cC}{{\cal C}}
\newcommand{\cD}{{\cal D}}

\newcommand{\cE}{{\cal E}}
\newcommand{\cP}{{\cal P}}

\newcommand{\bC}{\mbox{{\bf C}}}
\newcommand{\bZp}{\mbox{{\bf Z}}_{\geq 0}}
\newcommand{\bZ}{\mbox{{\bf Z}}}

\title
{
Dynamical correlation functions\\
of the spin Calogero-Sutherland model
}

\author
{%
Takashi Yamamoto%
\thanks{
CREST researcher:
Japan Science and Technology Corporation (JST),
Kawaguchi 332-0012, Japan.
e-mail:
yam@cmpt01.phys.tohoku.ac.jp}
{}  and{}
Mitsuhiro Arikawa%
\thanks{arikawa@cmpt01.phys.tohoku.ac.jp}
\\
Department of Physics, Tohoku University,
Sendai 980-8578,
Japan%
}

\begin{document}

\maketitle

\begin{abstract}
The thermodynamic limit
of the dynamical density and spin-density
two-point correlation functions
for the spin Calogero-Sutherland model
are derived
from Uglov's finite-size results.
The resultant formula
for the density two-point correlation function
is consistent with the previous conjecture
on the basis of the minimal number
of elementary excitations.
\end{abstract}

\newpage


\section{Introduction}
\label{introduction}
In recent years, a considerable number of studies
have been made on
the $SU(N)$ spin Calogero-Sutherland (CS) model
\cite{HaHaldane,MP2}.
This model describes $n$ particles system on a circle of length
$L$ interacting with the inverse-square type potential.
Each particle is labelled by its coordinate $x_j$ and spin
with $N(\geq 1)$ possible values.
(When $N=1$, this is the CS model \cite{Cal,Suth}.)
The Hamiltonian of the model is given by
\begin{equation}
\label{CShamiltonian}
H
=
-\frac{1}{2}\sum_{i=1}^n \frac{\partial^2}{\partial x_i^2}
+
\left(\frac{\pi}{L}\right)^2
\sum_{1\leq i<j\leq n}
\frac{\beta(\beta+P_{ij})}
     {\sin^2\frac{\pi}{L}(x_i-x_j)},
\end{equation}
where $\beta$ is a coupling parameter
and $P_{ij}$ is the spin exchange operator
for particles $i$ and $j$.
In this paper,
we take $N=2$ and $\beta$
to be a positive integer.

A lot of intriguing results have been obtained
in connection with the spin CS model.
In particular,
the eigenfunctions of the spin CS model
have been explicitly constructed and then,
using these properties,
the dynamical correlation functions of
this model were computed.
For $\beta=1$ which is the simplest nontrivial case,
the hole propergator of the $SU(2)$ spin CS model
(with finite $n$ and in the thermodynamic limit)
has been calculated by Kato \cite{Green}.
He also gave a conjectural formula for the arbitrary
integer coupling case.
Using the Jack polynomials with prescribed symmetry
\cite{BF1,Dunkl3},
this conjecture was recently confirmed by
Kato and one of the authors \cite{KaYa}.
On the other hand,
introducing the new class of orthogonal polynomials
\cite{TU,Uglov},
exact results have been obtained by Uglov \cite{Uglov}.
He computed the dynamical density and spin-density
two-point correlation functions of
the $SU(2)$ spin CS model with finite $n$.
In \cite{KYA},
Kato and authors have studied the construction for
the dynamical correlation functions of
the $SU(N)$ spin CS model
in the thermodynamic limit.
We gave the formula for
the density two-point correlation function
in the thermodynamic limit
and checked the consistency with the predictions
from conformal field theory \cite{Kawakami}.
However, the relation between Uglov's work and ours
is missing.

In this paper, from Uglov's formulae
for the dynamical correlation function
of the spin CS model,
we take the thermodynamic limit
which is technically nontrivial.
For the density two-point correlation function,
we prove our previous result in
the thermodynamic limit microscopically.

Here,
we give our main results together with those of
the work in ref. \cite{Green,KaYa}.
We introduce the following notations:
for non-negative integers $a$, $b$ and $c$,
\begin{eqnarray}
\label{ene-td}
\cE(u,v,w;a,b,c)
&=&
\sum_{i=1}^a\epsilon_{\rm p}(u_i)
+\sum_{j=1}^b\epsilon_{\rm h}(v_j)
+\sum_{k=1}^c\epsilon_{\rm h}(w_k),
\\
\label{mome-td}
\cP(u,v,w;a,b,c)
&=&
\frac{\pi\rho_0}{2}
\Big[
-(2\beta+1)\sum_{i=1}^a u_i
+\sum_{j=1}^b v_j
+\sum_{k=1}^c w_k
\Big],
\\
\label{int}
I(a,b,c)[\ast]
&=&
\prod_{i=1}^{a}
\int_1^\infty du_i
\prod_{j=1}^{b}
\int_{-1}^1dv_j
\prod_{k=1}^{c}
\int_{-1}^1dw_k
(\ast)
|F_\beta(u,v,w;a,b,c)|^2,
\end{eqnarray}
where
$\rho_0$ is the density of particles,
variables
$u=(u_1,\cdots,u_a)$,
$v=(v_1,\cdots,v_b)$ and $w=(w_1,\cdots,w_c)$ represent
the normalized momenta of quasiparticle with spin $\sigma$,
quasiholes with spin $-\sigma$ and $\sigma$,
respectively ($\sigma=\pm 1/2$).
The quasiparticle and quasihole dispersions
are introduced by
\ba
\label{particle-dis}
&&
\epsilon_{\rm p}(y)
=
(2\beta+1)^2
\frac{1}{2}
\left(\frac{\pi\rho_0}{2}\right)^2(y^2-1),
\\
\label{hole-dis}
&&
\epsilon_{\rm h}(y)
=
(2\beta+1)
\frac{1}{2}
\left(\frac{\pi\rho_0}{2}\right)^2(1-y^2),
\ea
respectively.
The function $F_\beta$ is defined by
\begin{eqnarray}
\label{form-factor}
&&F_\beta(u,v,w;a,b,c)
\nn
\\
&=&
\frac{
\prod_{1\leq i < j\leq b}
 (v_i - v_j )^{g_{\rm h}^{\rm d}}
\prod_{1\leq i < j\leq c}
 (w_i - w_j )^{g_{\rm h}^{\rm d}}
\prod_{i=1}^{b}\prod_{j=1}^{c}
 (v_i - w_j)^{g_{\rm h}^{\rm o}}
     }
     {
\prod_{i=1}^{a}\prod_{j=1}^{b}
(u_i - v_j)
\prod_{i=1}^{a}
(u_i^2-1)^{(1-g_{\rm p}^{\rm d})/2}
\prod_{j=1}^{b}
(1-v_j^2)^{(1-g_{\rm h}^{\rm d})/2}
\prod_{k=1}^{c}
(1-w_k^2)^{(1-g_{\rm h}^{\rm d})/2}
     },
\end{eqnarray}
where
\be
\label{stat-mat}
g_{\rm p}^{\rm d}=\beta+1,\
g_{\rm h}^{\rm d}=(\beta+1)/(2\beta+1),\
g_{\rm h}^{\rm o}=-\beta/(2\beta+1).
\ee

The retarded Green function \cite{Green,KaYa},
density and spin-density two-point correlation functions
can respectively be written as the following form:
\ba
\label{rGreen-td}
\langle \psi(x,t)\psi(0,0)\rangle
&=&
A(\beta)I(0,\beta+1,\beta)
[(\pi\rho_0/2)e^{{\rm i}(\cP(0,\beta+1,\beta)x
-(\cE(0,\beta+1,\beta)-\zeta)t)}],
\\
\label{dd-td}
\langle \rho(x,t)\rho(0,0)\rangle
&=&
B(\beta)I(1,\beta+1,\beta)
[\cP(1,\beta+1,\beta)^2
\cos(\cP(1,\beta+1,\beta)x)
e^{-{\rm i}\cE(1,\beta+1,\beta)t}],
\\
\label{ss-td}
\langle s(x,t)s(0,0)\rangle
&=&
C_{\mbox{\scriptsize I}}(\beta)I(1,\beta,\beta+1)
[(\pi\rho_0/2)^2\cos(\cP(1,\beta,\beta+1)x)
e^{-{\rm i}\cE(1,\beta,\beta+1)t}]
\nn
\\
&+&
C_{\mbox{\scriptsize II}}(\beta)I(1,\beta+2,\beta-1)
[(\pi\rho_0/2)^2\cos(\cP(1,\beta+2,\beta-1)x)
e^{-{\rm i}\cE(1,\beta+2,\beta-1)t}],
\ea
where
$\zeta=(\pi(2\beta+1)\rho_0/2)^2$ is
the chemical potential and
we use the convention
$\prod_{i=1}^0(\ast)$=1.
Constant factors in the above formulae are defined by
\ba
A(\beta)&=&
\frac{1}{\pi(2\beta+1)^\beta}D(\beta),
\\
B(\beta)&=&
\frac{1}{\pi^2(2\beta+1)^{\beta+1}}D(\beta),
\\
C_{\mbox{\scriptsize I}}(\beta)&=&
\frac{1}{4\pi^2(2\beta+1)^{\beta-1}}D(\beta),
\\
C_{\mbox{\scriptsize II}}(\beta)&=&
\frac{1}{4\pi^2(2\beta+1)^{\beta-1}}\frac{\beta}{\beta+2}D(\beta),
\ea
where, using the gamma function $\Gam(z)$,
the constant $D(\beta)$ is given by
\be
D(\beta)
=
\frac{1}{\Gam(\beta+2)}
\prod_{j=1}^{2\beta+1}
\frac{\Gam\big((\beta+1)/(2\beta+1)\big)}
     {\Gam^{2}\big(j/(2\beta+1)\big)}.
\ee

The paper is organized as follows.
In section \ref{Uglov-work},
we recall Uglov's results about
the dynamical correlation functions
of the $SU(2)$ spin CS model.
In section \ref{thermodynamic-limit},
firstly, we examine the excitation contents
of the intermediate states of
the dynamical correlation functions.
Secondly,
taking the thermodynamic limit,
we derive the formulae (\ref{dd-td}) and (\ref{ss-td}).
The conclusion is presented in section \ref{conclusion}.
Appendix A contains
proof of the statement in the subsection
\ref{intermediate-states}.
In appendix B,
we give the examples of the explicit
formulae of the building blocks
for the dynamical correlation functions.

\section{Uglov's formulae
for the dynamical correlation functions}
\label{Uglov-work}

In this section, we fix notations and then
recall Uglov's exact results
for the dynamical correlation functions \cite{Uglov}.
We only give the final results of Uglov's paper.
For details, see ref. \cite{Uglov}.

\subsection{Notations}
\label{notations}

In Uglov's formulation,
the states of the (transformed) Hamiltonian
are labeled by the colored partitions.
A brief mathematical preliminary here may be in order.
We fix notations which will be
to the fore in this paper (see refs. \cite{Macd,Uglov}).

For a fixed non-negative integer $n$, let $\Lambda_n
=\{\lam=(\lam_1,\lam_2,\cdots,\lam_n)\in(\bZp)^n\,
|\,\lam_1\geq\lam_2\geq\cdots\geq\lam_n\}$
be the set of all partitions with length less or equal to $n.$
The weight of a partition $\lam=(\lam_1,\lam_2,\cdots,\lam_n)$
is defined by
$|\lam |=\sum_{i=1}^n\lam_i$.
A partition can be represented by a Young diagram.
For example, the partition $\lam=(4,3,1)$ is expressed as
$$
\lam=
\Yvcentermath1
\yng(4,3,1)
$$
When there is a square in the $i$th row and $j$th
column of $\lam$, we write $(i,j)\in\lam$.
The conjugate of a partition $\lam=(\lam_1,\lam_2,\cdots,\lam_n)$
is the partition
$\lam'=(\lam'_1,\lam'_2,\cdots,\lam'_{\lam_1})$ whose diagram is the
transpose of the diagram $\lam$.
For instance, if $\lam=(4,3,1)$, then $\lam'=(3,2,2,1)$:
$$
\lam'=
\Yvcentermath1
\yng(3,2,2,1)
$$

Let $\lam=(\lam_1,\lam_2,\cdots,\lam_n)$ be a partition.
For a square $s=(i,j)\in\lam$, the numbers
\be
\begin{array}{ll}
\label{arm-leg}
a(s)=\lam_i-j,  \quad & a'(s)=j-1, \\
l(s)=\lam'_j-i, \quad & l'(s)=i-1,
\end{array}
\ee
are called arm-length, coarm-length, leg-length, and coleg-length,
respectively:
\newcommand{\sss}{\stackrel{{\tiny a}}{-}}
\newcommand{\arm}{a}
\newcommand{\coarm}{a'}
\newcommand{\leg}{l}
\newcommand{\coleg}{l'}
$$
\lam=
\Yvcentermath1
\young(%
\hfil\hfil\hfil\uparrow\hfil\hfil\hfil\hfil\hfil\hfil\hfil\hfil,%
\hfil\hfil\hfil\coleg\hfil\hfil\hfil\hfil\hfil\hfil\hfil,%
\hfil\hfil\hfil\downarrow\hfil\hfil\hfil\hfil\hfil\hfil,%
\leftarrow\coarm\rightarrow s\leftarrow-\arm-\rightarrow,%
\hfil\hfil\hfil\uparrow\hfil\hfil\hfil\hfil,%
\hfil\hfil\hfil |\hfil\hfil\hfil,%
\hfil\hfil\hfil\leg\hfil\hfil\hfil,%
\hfil\hfil\hfil |\hfil\hfil,%
\hfil\hfil\hfil\downarrow,%
\hfil\hfil
)
$$
Also the numbers
\ba
\label{content}
&& c(s)=a'(s)-l'(s)=j-i, \\
\label{hooklength}
&& h(s)=a(s)+l(s)+1=\lam_i+\lam'_j-i-j+1,
\ea
are called content and hook-length, respectively.
For $\alpha\in\bC$, their refinements are defined by
\ba
\label{modi-content}
&& c(s;\alpha)
=a'(s)-\alpha l'(s)=j-1-\alpha(i-1), \\
\label{upper-hook-length}
&& h_\lam^*(s;\alpha)
=a(s)+1+\alpha l(s)=\lam_i-j+1+\alpha(\lam'_j-i), \\
\label{lower-hook-length}
&& h^\lam_*(s;\alpha)
=a(s)+\alpha(l(s)+1)=\lam_i-j+\alpha(\lam'_j-i+1).
\ea
Moreover we define the following numbers:
\ba
&&
d(s;\alpha)
=h_\lam^*(s;\alpha)h^\lam_*(s;\alpha),\\
&&
e(s;\alpha)
=\frac{a'(s)+\alpha(n-l'(s))}{a'(s)+1+\alpha(n-l'(s)-1)}
=\frac{j-1+\alpha(n-i+1)}{j+\alpha(n-i)}.
\ea

We recall a coloring scheme of diagrams.
(Here we only need a coloring by two colors, white and black,
since we consider the case with $N=2$.)
For a partition $\lam$, we define two subsets of $\lam$ by
$W_\lam=\{s\in\lam\,|\,c(s)\equiv 0\ \mbox{mod}\, 2\}$
and
$B_\lam=\{s\in\lam\,|\,c(s)\equiv 1\ \mbox{mod}\, 2\}$.
We call the color of $s\in\lam$ white (black)
if $s\in W_\lam$ ($\in B_\lam$),
and call $\lam=W_\lam\sqcup B_\lam$ the colored partition.
(Notice that $(1,1)\in W_\lam$ (if $\lam\ne\O$).)
For example, if $\lam=(4,3,1)$, then
$W_\lam=\{\Yvcentermath1\young(\hfill)\  \mbox{in the following diagram}\}$,
and
$B_\lam=\{\Yvcentermath1\young(\bullet)\  \mbox{in the following
diagram}\}$:
$$
\lam=
\Yvcentermath1
\young(
\hfill\bullet\hfil\bullet,%
\bullet\hfil\bullet,%
\hfil
)
$$

We define an another subset of $\lam$ by
$H_2(\lam)=\{s\in\lam\,|\,h(s)\equiv 0\ \mbox{mod}\, 2\}$.
For example, if $\lam=(4,3,1)$, then
$H_2(\lam)=\{\Yvcentermath1\young(\star)\  \mbox{in the following
diagram}\}$:
$$
\lam=
\Yvcentermath1
\young(
\star\star\hfil\hfil,%
\star\star\hfil,%
\hfil
)
$$

\subsection{Dynamical correlation functions}
\label{dcf-finite}

We now recall Uglov's formulae
for the dynamical correlation functions \cite{Uglov}.
We take the number of particles $n$
to be an even number such that $n/2$ is odd \cite{Uglov}.

In Uglov's formalism,
the states of the CS Hamiltonian
are labeled by the colored partitions.
First of all, the total energy
with respect to the transformed Hamiltonian,
total momentum, and
total $z$-component of spin for the colored partition
$\lam$ are respectively given by\footnote{%
The formula for $E_\lam$ in ref. \cite{Uglov}
has typographical error.}
\ba
\label{ene-finite}
&&
E_\lam
=
\frac{1}{2}\left(\frac{2\pi}{L}\right)^2
\left[
n_w(\lam')-\gam n_w(\lam)
+
\frac{1}{2}((n-1)\gam+1)|W_\lam|
\right],                 \\
\label{mome-finite}
&&
P_\lam
=
\frac{2\pi}{L}|W_\lam|, \\
\label{spin-finite}
&&
S_\lam
=
|W_\lam|-|B_\lam|,
\ea
where $\gam=2\beta+1(\in\bZ_{>0})$, and
\ba
&&
n_w(\lam)=\sum_{s\in W_\lam}l'(s), \\
&&
n_w(\lam')=\sum_{s\in W_\lam}a'(s).
\ea
Here, for any subset $\mu\subset\lam$, we denote by $|\mu|$
the number of squares in $\mu$.

Next, the building blocks for the main factors of
the dynamical correlation functions are defined as follows:
for a colored partition $\lam\in\Lambda_n$,
\ba
&&
X_\lam
=\prod_{s\in W_\lam\setminus\{(1,1)\}} c(s;\gam)^2,
\\
&&
Y_\lam
=\prod_{s\in H_2(\lam)} d(s;\gam),
\\
&&
Z_\lam
=\prod_{s\in W_\lam} e(s;\gam).
\ea

For the system with Hamiltonian (\ref{CShamiltonian}),
we denote the ground state expectation value
of the operator ${\cal O}$ by $\langle{\cal O}\rangle_n$.
Then,
the (ground state) dynamical density
and spin-density two point correlation functions
are respectively given by \cite{Uglov}
\ba
\label{dd-finite}
\langle\rho(x,t)\rho(0,0)\rangle_n
&=&
\frac{4}{\pi^2}
\sum_{{{{{\footnotesize \lam\in\Lambda_n:\,
                        \mbox{{\scriptsize colored partition}}}}
          \atop
         {\footnotesize |\lam|:\,\mbox{{\scriptsize even}}}}
          \atop
         {\footnotesize S_\lam=0}}
          \atop
         {\footnotesize |W_\lam|=|H_2(\lam)|}}
|P_\lam|^2 X_\lam Y_\lam^{-1} Z_\lam
e^{-itE_\lam}\cos(xP_\lam),
\\
\label{ss-finite}
\langle s(x,t)s(0,0)\rangle_n
&=&
\frac{1}{2L^2}
\sum_{{{{{\footnotesize \lam\in\Lambda_n:\,
                        \mbox{{\scriptsize colored partition}}}}
          \atop
         {\footnotesize |\lam|:\, \mbox{{\scriptsize odd}}}}
          \atop
         {\footnotesize S_\lam=\pm 1}}
          \atop
         {\footnotesize |W_\lam|=|H_2(\lam)|+1}}
X_\lam Y_\lam^{-1} Z_\lam
e^{-itE_\lam}\cos(xP_\lam),
\ea
where $\rho(x,t)$ and $s(x,t)$
are the Heisenberg representations of
the reduced density operator
$\rho(x)=\sum_{i=1}^n\delta(x-x_i)-n/L$
and the $z$-component of spin-density operator
$s(x)=\sum_{i=1}^n\delta(x-x_i)\sigma_i^z/2$,
respectively.
Here $\sigma_i^z$ is the $z$-component of Pauli matrices.

\section{Thermodynamic limit of
the dynamical correlation functions}
\label{thermodynamic-limit}

In this section,
we take the thermodynamic limit of Uglov's exact
formulae (\ref{dd-finite}) and (\ref{ss-finite}).

Firstly,
we determine the excitation contents
of the intermediate states
for the dynamical correlation functions
(\ref{dd-finite}) and (\ref{ss-finite}).
Secondly,
we rewrite the formulae (\ref{dd-finite}) and (\ref{ss-finite})
in terms of parameters which
correspond to the elementary excitations.
Finally, we take the thermodynamic limit.

\subsection{Intermediate states}
\label{intermediate-states}

In order to take the thermodynamic limit,
we must determine the excitation contents
of the intermediate states
for the dynamical correlation functions
(\ref{dd-finite}) and (\ref{ss-finite}).

Except for the factor $X_\lam$
which comes from the matrix element
of the local operators,
the factors in the sums of
the right hand side of (\ref{dd-finite}) and (\ref{ss-finite})
are non-zero for all $\lam$.
Therefore, a summand in the sums of
the right hand side of (\ref{dd-finite}) and (\ref{ss-finite})
is non-zero if and only if $X_\lam\ne 0$.
From the definition, it is easy to see that
$X_\lam\ne 0\Leftrightarrow (2,\gam+1)\notin\lam$ \cite{Uglov}.
(Notice that $\gam=2\beta+1\in\bZ_{>0}$.)
As is the spinless case \cite{Ha,LPS},
this condition has the following physical interpretation:
The intermediate states contributing to
the dynamical correlation functions
(\ref{dd-finite}) and (\ref{ss-finite})
have one quasiparticle and $\gam$ quasihole excitations.
We will see that these elementary excitations
have the $SU(2)$ spin degrees of freedom.
The above condition together with the conditions
on the sums in the formulae
(\ref{dd-finite}) and (\ref{ss-finite})
determine the intermediate states.

We parametrize the intermediate states
of the correlation functions
(\ref{dd-finite}) and (\ref{ss-finite}).
For this purpose,
we introduce some notations.
In consideration of above observation,
we define the subset of $\Lambda_n$ by
$\Lambda_n^{(\gam)}
=\{\lambda\in\Lambda_n\,|\,(2,\gam+1)\notin\lam\}$.
That is, a partition $\lam\in\Lambda_n^{(\gam)}$
has $\gam$ columns and one `arm'.
For a partition
$\lam=(\lam_1,\lam_2,\cdots,\lam_n)\in\Lambda_n^{(\gam)}$,
we introduce the notation
$
\lam
=
\langle
\lam'_1,\lam'_2,\cdots,\lam'_\gam;r
\rangle,
$
which consists of $\gam$ columns and one `arm'.
Here $r=\lam_1-\gam$. (If $\lam_1<\gam$, then $r=0$.)
For example,
$\lam=(13,5,5,5,4,4,4,2,2,1)=\langle 10,9,7,7,4;8\rangle\in\Lambda_n^{(5)}$:
$$
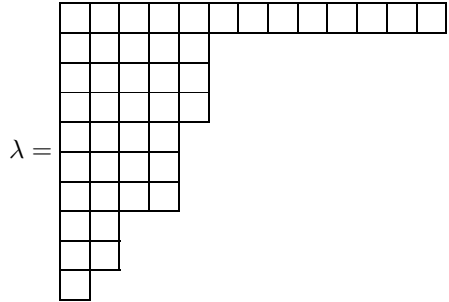
\lam=
\Yvcentermath1
\yng(13,5,5,5,4,4,4,2,2,1)
$$

Using these notations, we can state that
the intermediate states
for the density two-point
correlation function (\ref{dd-finite})
are colored partitions $\lam\in\Lambda_n^{(\gam)}$
with even weight, $S_\lam=0$ and $|W_\lam|=|H_2(\lam)|$.
We call these colored partitions the d-d admissible.
Similarly,
the intermediate states
for the spin-density two-point
correlation function (\ref{ss-finite})
are colored partitions $\lam\in\Lambda_n^{(\gam)}$
with odd weight, $S_\lam=\pm 1$ and $|W_\lam|=|H_2(\lam)|+1$.
We call these colored partitions the s-s admissible.

The above conditions on the intermediate states
of dynamical correlation functions
are rather complicated.
Technical difficulty in taking
the thermodynamic limit
comes form these complicated conditions.
Then, in the following, we simplify the conditions
for d-d and s-s admissible colored partitions.
For this purpose, we introduce more notations.

For
$\nu
=(\nu_1,\cdots,\nu_{\gam+1})\in \{0,1\}^{\times(\gam+1)}$,
we define two subsets $I_1(\nu)$ and $I_2(\nu)$
of $I=\{1,\cdots,\gam\}$ by
\ba
I_1(\nu)
&=&
\left\{j\in\{1,\cdots,\gam\} \,\left|\,
\nu_j
=
\left\{
\begin{array}{ll}
(1+(-1)^j)/2,\quad     & \mbox{if } \nu_{\gam+1}=0 \\
(1+(-1)^{j-1})/2,\quad & \mbox{if } \nu_{\gam+1}=1
\end{array}
\right.
\right.
\right\},
\\
I_2(\nu)
&=&
I\setminus I_1(\nu).
\ea
For example,
if
$\nu
=(\overbrace{1,1,\cdots,1}^{\gam+1\ \mbox{{\footnotesize times}}})$,
then
$I_1(\nu)=\{j\in I \,|\, j : \mbox{odd}\}$
and
$I_2(\nu)=\{j\in I \,|\, j : \mbox{even}\}$.
We introduce the function
$
\label{parity}
\rho:\ \Lambda_n^{(\gam)}
  \longrightarrow\{0,1\}^{\times(\gam+1)}
$
by
$$
\rho(\lam)
=(\sigma(\lam'_1),\cdots,\sigma(\lam'_\gam),\sigma(r)),\,
\mbox{if}\,\,
\lam
=
\langle
\lam'_1,\cdots,\lam'_\gam;r
\rangle,
$$
where
$\sigma(a)=0\ (1)$ if $a$ is even (odd).
We call $\rho(\lam)$ the parity of $\lam\in\Lambda_n^{(\gam)}$.

It can be easy to show that,
for a colored partition
$\lam\in\Lambda_n^{(\gam)}$ with parity
$\rho(\lam)=(\nu_1,\cdots,\nu_{\gam+1})$,
\ba
&&
\label{w-b}
S_\lam
=|W_\lam|-|B_\lam|=\sum_{j=1}^{\gam+1}(-1)^{j-1}\nu_j.
\ea
Using these formulae, we have
\be
\label{cond1}
S_\lam
=
\left\{
\begin{array}{rr}
0 \\
\pm 1
\end{array}
\right.
\Leftrightarrow
\#I_1(\rho(\lam))
=
\left\{
\begin{array}{ll}
\beta+1 \\
\beta\,\mbox{or}\,\beta+2.
\end{array}
\right.
\ee
Here, for a set $A$, $\#A$ denotes the number of elements.
Moreover we can show that,
for $\lam\in\Lambda_n$ with even (resp. odd) weight,
\be
\label{cond2}
S_\lam=0\quad(\mbox{resp. } \pm 1)
\Leftrightarrow
|W_\lam|=|H_{2}(\lam)|\quad
(\mbox{resp. } |H_{2}(\lam)|+1).
\ee
The proof of the statement (\ref{cond2})
is given in Appendix A.
The statements (\ref{cond1}) and (\ref{cond2})
are essential to taking the thermodynamic limit.

From above statements,
we see that a colored partition
$\lam\in\Lambda_n^{(\gam)}$
is the d-d admissible
if and only if
$|\lam|$ is even and $\#I_1(\rho(\lam))=\beta+1$.
Similarly,
a colored partition
$\lam\in\Lambda_n^{(\gam)}$
is the s-s admissible
if and only if $|\lam|$ is odd and
$\#I_1(\rho(\lam))=\beta$ or $\beta+2$.
The s-s admissible colored partitions are divided into
two types which are characterized by
$\#I_1(\rho(\lam))=\beta$ or $\beta+2$.
We call former type I and latter type II.

The $SU(2)$ spin degrees of freedom of the
elementary excitations
are assigned as follows.
For an admissible colored partition
$\lam=
\langle
\lam'_1,\cdots,\lam'_\gam;r
\rangle\in\Lambda_n^{(\gam)}$ with
parity $\rho(\lam)=(\nu_1,\cdots,\nu_{\gam+1})$,
the spin of quasiparticle is $1/2$ (resp. $-1/2$) if
$\nu_{\gam+1}=0$ (resp. $1$).
On the other hand,
the spin of quasihole corresponding to $\lam'_j$
is $1/2$ (resp. $-1/2$)
if $\nu_{\gam+1}=0$ and $j\in I_2(\rho(\lam))$
or $\nu_{\gam+1}=1$ and $j\in I_1(\rho(\lam))$
(resp.
$\nu_{\gam+1}=0$ and $j\in I_1(\rho(\lam))$
or $\nu_{\gam+1}=1$ and $j\in I_2(\rho(\lam))$
).

Now, we can determine the excitation contents
of the intermediate states
for the dynamical correlation functions.
The excitation contents of the intermediate states
for the dynamical density two-point correlation function
is given by the following set of
quasiparticle and quasiholes:
\be
\label{minimal-bubble-dd}
\left\{
\begin{array}{l}
\mbox{one quasiparticle with spin $\sigma$}
\\
\mbox{$\beta+1$ quasiholes with spin $-\sigma$}
\\
\mbox{$\beta$ quasiholes with spin $\sigma$},
\end{array}
\right.
\ee
where $\sigma=\pm 1/2$.
This is consistent with the result in ref. \cite{KYA}.
Similarly,
the excitation contents of the intermediate states
for the dynamical spin-density two-point correlation function
is given by the following sets of
quasiparticle and quasiholes:
\be
\label{minimal-bubble-ss1}
\left\{
\begin{array}{l}
\mbox{one quasiparticle with spin $\sigma$}
\\
\mbox{$\beta$ quasiholes with spin $-\sigma$}
\\
\mbox{$\beta+1$ quasiholes with spin $\sigma$},
\end{array}
\right.
\ee
and
\be
\label{minimal-bubble-ss2}
\left\{
\begin{array}{l}
\mbox{one quasiparticle with spin $\sigma$}
\\
\mbox{$\beta+2$ quasiholes with spin $-\sigma$}
\\
\mbox{$\beta-1$ quasiholes with spin $\sigma$}.
\end{array}
\right.
\ee
It is remarkable that
two types of the set
of the elementary excitations contribute to
the dynamical spin-density two-point correlation function.

For later convenience, using (\ref{cond1}) and (\ref{cond2}),
we rewrite (\ref{dd-finite}) and (\ref{ss-finite}) as
\ba
\label{dd-finite-parity}
\langle\rho(x,t)\rho(0,0)\rangle_n
&=&
\frac{4}{\pi^2}
\sum_{\nu=(\nu_1,\cdots,\nu_{\gam+1})\in \{0,1\}^{\times(\gam+1)}
       \atop
         {\scriptscriptstyle \#I_1(\nu)=\beta+1}}
\sum_{{{{{\footnotesize \lam\in\Lambda_n^{(\gam)}:\,
                        \mbox{{\scriptsize colored partition}}}}
          \atop
         {\footnotesize |\lam|:\,\mbox{{\scriptsize even}}}}
          \atop
         {\footnotesize \rho(\lam)=\nu}}
     }
|P_\lam|^2 X_\lam Y_\lam^{-1} Z_\lam
e^{-itE_\lam}\cos(xP_\lam),
\\
\label{ss-finite-parity}
\langle s(x,t)s(0,0)\rangle_n
&=&
\frac{1}{2L^2}
\sum_{\nu=(\nu_1,\cdots,\nu_{\gam+1})\in \{0,1\}^{\times(\gam+1)}
       \atop
         {\scriptscriptstyle \#I_1(\nu)=\beta}}
\sum_{{{{{\footnotesize \lam\in\Lambda_n^{(\gam)}:\,
                        \mbox{{\scriptsize colored partition}}}}
          \atop
         {\footnotesize |\lam|:\,\mbox{{\scriptsize odd}}}}
          \atop
         {\footnotesize \rho(\lam)=\nu}}
     }
X_\lam Y_\lam^{-1} Z_\lam
e^{-itE_\lam}\cos(xP_\lam)
\nn
\\
&+&
\frac{1}{2L^2}
\sum_{\nu=(\nu_1,\cdots,\nu_{\gam+1})\in \{0,1\}^{\times(\gam+1)}
       \atop
         {\scriptscriptstyle \#I_1(\nu)=\beta+2}}
\sum_{{{{{\footnotesize \lam\in\Lambda_n^{(\gam)}:\,
                        \mbox{{\scriptsize colored partition}}}}
          \atop
         {\footnotesize |\lam|:\,\mbox{{\scriptsize odd}}}}
          \atop
         {\footnotesize \rho(\lam)=\nu}}
     }
X_\lam Y_\lam^{-1} Z_\lam
e^{-itE_\lam}\cos(xP_\lam).
\ea
The first summation in the right hand side
of (\ref{dd-finite-parity}) is
taken over $2{}_\gam C_{(\gam+1)/2}$ different parities, since
\be
2{}_\gam C_{(\gam+1)/2}
=
\#\{
\nu
=(\nu_1,\cdots,\nu_{\gam+1})\in \{0,1\}^{\times(\gam+1)}\,|\,
\#I_1(\nu)=\beta+1
\}.
\ee
Similarly, the first summations of the first and second line
in the right hand side of (\ref{ss-finite-parity}) are
respectively taken over $2{}_\gam C_{(\gam-1)/2}$ and
$2{}_\gam C_{(\gam+3)/2}$ different parities.

\subsection{Quasiparticle and quasihole description
of the dynamical correlation functions}
\label{quasiparticle-description}

To take the thermodynamic limit,
next our task is to rewrite the formulae
(\ref{dd-finite-parity}) and (\ref{ss-finite-parity})
in terms of parameters which
correspond to the momenta
of quasiholes and quasiparticle
(see \cite{Ha,LPS}).
We have already introduced such parameters,
{\it i.e.},
$
\lam
=
\langle
\lam'_1,\lam'_2,\cdots,\lam'_\gam;r
\rangle.
$
The quantities $\lam'_1,\lam'_2,\cdots,\lam'_\gam$ and $r$
are respectively related to the momenta
of quasiholes and quasiparticle.

Although it can be possible to proceed the calculation
by using the parameters
$\lam'_1,\lam'_2,\cdots,\lam'_\gam$ and $r$,
it is appropriate to introduce new parameters as follows.
We define the following numbers \cite{Uglov}:
\ba
\label{row-w}
&&w_i(\lam)
=
\#\{s\in i\mbox{th row of }\lam\,|\,s: \mbox{white}\},
\\
\label{column-w}
&&w_j(\lam')
=
\#\{s\in j\mbox{th column of }\lam\,|\,s: \mbox{white}\}.
\ea
We note that,
using these numbers, we have
$|W_\lam|=\sum_{i=1}^n w_i(\lam)
=\sum_{j=1}^{\lam_1} w_j(\lam')$.
Then,
instead of
$\lam'_1,\lam'_2,\cdots,\lam'_\gam$ and $r$,
we adopt
$w_1(\lam'),\cdots,w_\gam(\lam')$, and $p=w_1(\lam)-\gam$
as the parameters.
These two sets of parameters are related by the formulae
\be
\lam'_j
=
\left\{
\begin{array}{ll}
2w_j(\lam')-1,\quad & \mbox{if $j$: odd, $\lam'_j$: odd}, \\
2w_j(\lam'),  \quad & \mbox{if $\lam'_j$: even},\\
2w_j(\lam')+1,\quad & \mbox{if $j$: even, $\lam'_j$: odd},
\end{array}
\right.
\ee
and
\be
r
=
\left\{
\begin{array}{ll}
2p+1,\quad & \mbox{if $r$: odd}, \\
2p,  \quad & \mbox{if $r$: even}.
\end{array}
\right.
\ee

Let us rewrite the formulae for the dynamical
correlation functions (\ref{dd-finite-parity}) and
(\ref{ss-finite-parity})
by using
$w_1(\lam'),\cdots,w_\gam(\lam')$, and $p$.
First of all,
for $\lam\in\Lambda_n^{(\gam)}$,
we have
\ba
&&
|W_\lam|
=\sum_{j=1}^{\lam_1} w_j(\lam')
=\sum_{j=1}^{\gam} w_j(\lam')+p, \\
&&
n_w(\lam)=\sum_{s\in W_\lam}l'(s)
=\sum_{{\scriptstyle 1\leq j\leq\gam}
       \atop
       {\scriptstyle j:\, \mbox{{\scriptsize odd}}}}
(w_j(\lam')^2-w_j(\lam'))
+\sum_{{\scriptstyle 2\leq j\leq\gam-1}
       \atop
     {\scriptstyle j:\, \mbox{{\scriptsize even}}}}
w_j(\lam')^2,            \\
&&
n_w(\lam')=\sum_{s\in W_\lam}a'(s)
=
\sum_{j=1}^{\gam}(j-1)w_j(\lam')+p(p+\gam).
\ea
Then, from the definitions
(\ref{ene-finite}) and (\ref{mome-finite}),
we obtain the formulae for $E_\lam$ and $P_\lam$.

Next we rewrite $X_\lam$, $Y_\lam$ and $Z_\lam$.
For this purpose, following Ha \cite{Ha},
we decompose a partition
$\lam=(\lam_1,\lam_2,\cdots,\lam_n)\in\Lambda_n^{(\gam)}$
into three sub-diagrams
$\lam=\cA_\lam\sqcup\cB_\lam\sqcup\cC_\lam$ where
\ba
&&
\cA_\lam=\{(1,j)\in\Lambda_n^{(\gam)}\,|\,1\leq j\leq\gam\},\\
&&
\cB_\lam=\{(i,j)\in\Lambda_n^{(\gam)}\,|\,1\leq j\leq\gam, 2\leq
i\leq\lam'_j\},\\
&&
\cC_\lam=\{(1,j)\in\Lambda_n^{(\gam)}\,|\,\gam+1\leq j\leq\lam_1\}.
\ea
For example, if
$\lam=(13,5,5,5,4,4,4,2,2,1)=\langle 10,9,7,7,4;8\rangle\in\Lambda_n^{(5)}$,
then
$\cA_\lam=$
sub-diagram which contains
$\Yvcentermath1\young(\heartsuit)$
in the following diagram,
$\cB_\lam=$
sub-diagram which contains
$\Yvcentermath1\young(\spadesuit)$,
and
$\cC_\lam=$
sub-diagram which contains
$\Yvcentermath1\young(\clubsuit)$:
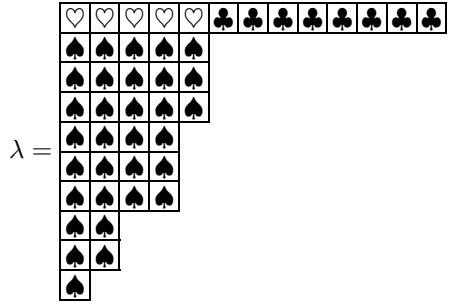
$$
\lam=
\Yvcentermath1
\young(
\heartsuit\heartsuit\heartsuit\heartsuit\heartsuit
\clubsuit\clubsuit\clubsuit\clubsuit\clubsuit\clubsuit\clubsuit\clubsuit,%
\spadesuit\spadesuit\spadesuit\spadesuit\spadesuit,%
\spadesuit\spadesuit\spadesuit\spadesuit\spadesuit,%
\spadesuit\spadesuit\spadesuit\spadesuit\spadesuit,%
\spadesuit\spadesuit\spadesuit\spadesuit,%
\spadesuit\spadesuit\spadesuit\spadesuit,%
\spadesuit\spadesuit\spadesuit\spadesuit,%
\spadesuit\spadesuit,%
\spadesuit\spadesuit,%
\spadesuit
)
$$
We denote
$W_{\cD_\lam}=W_\lam\cap\cD_\lam$,
$B_{\cD_\lam}=B_\lam\cap\cD_\lam$
and
$H_{2, \cD_\lam}=H_2(\lam)\cap\cD_\lam$ for $\cD=\cA,\cB,\cC$.
(Notice that $(1,1)\in W_{\cA_\lam}$ (if $\lam\ne\O$).)

For a colored partition $\lam\in\Lambda_n^{(\gam)}$,
we denote
$X_{\cD_\lam}
=\prod_{s\in W_{\cD_\lam}\setminus\{(1,1)\}} c(s;\gam)^2$ and
$Y_{\cD_\lam}
=\prod_{s\in H_{2, \cD_\lam}} d(s;\gam)$
for $\cD=\cA,\cB,\cC$.
It is obvious that
$X_\lam=X_{\cA_\lam} X_{\cB_\lam} X_{\cC_\lam}$, $Y_\lam=Y_{\cA_\lam}
Y_{\cB_\lam} Y_{\cC_\lam}$ and
$Z_\lam=Z_{\cA_\lam} Z_{\cB_\lam} Z_{\cC_\lam}$.
Then, it is easy to show that, for $\lam\in\Lambda_n^{(\gam)}$,
\ba
&&
X_{\cA_\lam}
=
2^{\gam-1}\Gam^2\big((\gam+1)/2\big),\\
&&
X_{\cB_\lam}
=
\xi^{\gam+1}
\prod_{j=1}^\gam
\xi^{-2w_j(\lam')}
\Gam^{-2}(j/\gam)
\nn\\
&&
\times
\prod_{{\scriptstyle 1\leq j\leq\gam}
       \atop
       {\scriptstyle j:\, \mbox{{\scriptsize odd}}}}
\Gam^2\big(w_j(\lam')-\xi(j-1)\big)
\prod_{{\scriptstyle 2\leq j\leq\gam-1}
       \atop
       {\scriptstyle j:\, \mbox{{\scriptsize even}}}}
\Gam^2\big(w_j(\lam')+1/2-\xi(j-1)\big),\\
&&
X_{\cC_\lam}
=
2^{2p}
\frac{\Gam^2\big(p+(\gam+1)/2\big)}
     {\Gam^2\big((\gam+1)/2\big)}, \\
&&
Y_{\cC_\lam}
=
2^{2p}
\Gam\big(p+1\big)
\frac{\Gam\big(p+(\gam+1)/2\big)}
     {\Gam\big((\gam+1)/2\big)},\\
&&
Z_{\cA_\lam}
=
\prod_{{\scriptstyle 1\leq j\leq\gam}
       \atop
       {\scriptstyle j:\, \mbox{{\scriptsize odd}}}}
\frac{\gam n+j-1}{\gam n+j-\gam},\\
&&
Z_{\cB_\lam}
=
\prod_{{\scriptstyle 1\leq j\leq\gam}
       \atop
       {\scriptstyle j:\, \mbox{{\scriptsize odd}}}}
\frac{\Gam\big(n/2+\xi j-\xi\big)\Gam\big(n/2-w_j(\lam')+\xi j+1/2\big)}
     {\Gam\big(n/2+\xi j-1/2\big)\Gam\big(n/2-w_j(\lam')+\xi j-\xi+1\big)}
\nn\\
&&\ \ \ \
\times
\prod_{{\scriptstyle 2\leq j\leq\gam-1}
       \atop
       {\scriptstyle j:\, \mbox{{\scriptsize even}}}}
\frac{\Gam\big(n/2+\xi j-\xi+1/2\big)\Gam\big(n/2-w_j(\lam')+\xi j\big)}
     {\Gam\big(n/2+\xi j\big)\Gam\big(n/2-w_j(\lam')+\xi j-\xi+1/2\big)},\\
&&
Z_{\cC_\lam}
=
\frac{\Gam\big(\gam n/2+1\big)\Gam\big(\gam n/2+p+(\gam+1)/2\big)}
     {\Gam\big(\gam n/2+(\gam+1)/2\big)\Gam\big(\gam n/2+p+1\big)},
\ea
where  $\xi=(2\gam)^{-1}$.
Notice that we can derive
all above formulae without fixing
the parity $\rho(\lam)$.

On the other hand, to derive
the explicit forms of $Y_{\cA_\lam}$ and $Y_{\cB_\lam}$
for $\lam\in\Lambda_n^{(\gam)}$, we must fix
the parity $\rho(\lam)$.
In fact, to write down the explicit forms
of $Y_{\cA_\lam}$ and $Y_{\cB_\lam}$,
we need more complicated notations.
However, for the purpose of taking the thermodynamic limit,
the necessary information are the sets $I_1(\rho(\lam))$,
$I_2(\rho(\lam))$, and the quantities of order
${\cal O}(n)$.
We see that, after replacing
the elements of sets $I_1$ and $I_2$ appropriately,
the thermodynamic limit of $Y_{\cD_\lam}$ and $Y_{\cD_{\lam'}}$
with $\rho(\lam)\ne\rho(\lam')$
coincide with each other $(\cD=\cA, \cB)$.
We do not give the explicit forms of
$Y_{\cA_\lam}$ and $Y_{\cB_\lam}$ for
the general admissible colored partition $\lam$.
In Appendix B, we give the examples
for some admissible colored partitions.
The thermodynamic limit of $Y_{\cA_\lam}$ and
$Y_{\cB_\lam}$ for general admissible colored partitions
are easily obtained from those examples.

Finally, we change the summation indices for
the sums in the dynamical correlation functions.
For example, we rewrite the sum
in the density two-point
correlation function (\ref{dd-finite-parity})
as
\be
\label{sum-dd}
\sum_{\nu=(\nu_1,\cdots,\nu_{\gam+1})\in \{0,1\}^{\times(\gam+1)}
       \atop
         {\scriptscriptstyle \#I_1(\nu)=\beta+1}}
\quad
\sum_{p\geq 0}
\quad
\sum_{n/2\geq w_{j_1}(\lam')\geq\cdots\geq w_{j_{\beta+1}}(\lam')\geq 0}
\quad
\sum_{n/2\geq w_{k_1}(\lam')\geq\cdots\geq w_{k_{\beta}}(\lam')\geq 0},
\ee
where $\{j_l\}_{l=1}^{\beta+1}=I_1(\rho(\lam)=\nu)$
such that
$j_1\geq\cdots\geq j_{\beta+1}$
and
$\{k_l\}_{l=1}^{\beta}=I_2(\rho(\lam)=\nu)$
with
$k_1\geq\cdots\geq k_{\beta}$.

\subsection{Thermodynamic limit}
\label{limit}

In this subsection,
we take the thermodynamic limit, {\it i.e.},
$n\rightarrow\infty$, $L\rightarrow\infty$
with $\rho_0=n/L$ fixed.

Let us introduce the momenta
$u$ and $v_j$ for $j=1,\cdots,\gam$ of
the quasiparticle and quasiholes, respectively,
by the formulae,
\ba
\label{particle-velocity}
\frac{1}{\gam}\frac{p}{n}
&\longrightarrow&
-\frac{u+1}{4},
\\
\label{hole-velocity}
\frac{w_j(\lam')}{n}
&\longrightarrow&
\frac{v_j+1}{4}.
\ea
Then we have the thermodynamic limit
of the energy and total momentum,
\ba
\label{ene-td'}
&&
E_\lam
\longrightarrow
\cE
=
\sum_{j=1}^\gam\epsilon_{\rm h}(v_j)+\epsilon_{\rm p}(u),
\\
\label{mome-td'}
&&
P_\lam
\longrightarrow
\cP
=
\frac{\pi\rho_0}{2}
\Big[
\sum_{j=1}^\gam v_j-\gam u
\Big],
\ea
where
\ba
\label{hole-dis'}
&&
\epsilon_{\rm h}(y)
=
\gam
\frac{1}{2}
\left(\frac{\pi\rho_0}{2}\right)^2(1-y^2), \\
\label{particle-dis'}
&&
\epsilon_{\rm p}(y)
=
\gam^2
\frac{1}{2}
\left(\frac{\pi\rho_0}{2}\right)^2(y^2-1).
\ea
We have adopted the normalization
(\ref{particle-velocity}) and (\ref{hole-velocity})
of $u$ and $v_j$ so that
the Fermi points coincides with $\{\pm 1\}$.

Also, using the formula
$\lim_{|a|\rightarrow \infty}\Gam(a+z)/\Gam(a)=a^z$,
we can obtain the thermodynamic limit of
$X_{\lam}Y_{\lam}^{-1}Z_{\lam}$.
In the following,
we consider the case of density two-point
correlation function.
In this case, we have
\ba
&&
X_{\lam}Y_{\lam}^{-1}Z_{\lam}
\longrightarrow
L^{-(\gam+1)}2^{2\gam}(\gam\rho_0)^{-(\gam+1)}
\Gam\big((\gam+1)/2\big)
\prod_{j=1}^\gam
\Gam\big(\xi+1/2\big)
\Gam^{-2}\big(j/\gam\big)
\nn
\\
&&
\qquad
\times
(u^2-1)^{(\gam-1)/2}
\prod_{j=1}^\gam
(1-v_j^2)^{\xi-1/2}
\prod_{j\in I_1(\rho(\lam))}
(u-v_j)^{-2}
\nn
\\
&&
\qquad
\times
\prod_{s=1,2}
\prod_{{\scriptstyle j, k\in I_s(\rho(\lam))}
       \atop
       {\scriptstyle j<k}}
(v_j-v_k)^{-2(\xi+1/2)}
\prod_{j\in I_1(\rho(\lam))}
\prod_{k\in I_2(\rho(\lam))}
(v_j-v_k)^{-2(\xi-1/2)}
\ea
for a d-d admissible colored partition
$\lam\in\Lambda_n^{(\gam)}$ with the fixed parity
$\rho(\lam)$.

For each d-d admissible colored partition
$\lam\in\Lambda_n^{(\gam)}$ with the fixed parity
$\rho(\lam)$,
we replace $\{v_j\}_{j\in I_1(\rho(\lam))}$
and $\{v_j\}_{j\in I_2(\rho(\lam))}$
by $\{v_j\}_{j=1}^{\beta+1}$
such that $v_1\geq \cdots\geq v_{\beta+1}$
and $\{w_j\}_{j=1}^\beta$
such that $w_1\geq \cdots\geq w_{\beta}$,
respectively.

In the thermodynamic limit,
we rewrite the sums as integrals
\ba
\label{sum2int-hole}
&&
\sum_{n/2\geq w_{j_1}(\lam')\geq\cdots\geq w_{j_{\beta+1}}(\lam')\geq 0}
\quad
\sum_{n/2\geq w_{k_1}(\lam')\geq\cdots\geq w_{k_{\beta}}(\lam')\geq 0}
\nn
\\
&&
\longrightarrow
L^\gam2^{-2\gam}\rho_0^\gam
\int_{1\geq v_{1}\geq v_{2}\geq\cdots\geq v_{\beta+1}\geq -1}
dv_1dv_2\cdots dv_{\beta+1}
\int_{1\geq w_{1}\geq w_{2}\geq\cdots\geq w_{\beta}\geq -1}
dw_1dw_2\cdots dw_{\beta},
\\
\label{sum2int-particle}
&&
\sum_{p\geq 0}
\longrightarrow
-
L2^{-2}\gam\rho_0\int_{-\infty}^{-1}du.
\ea

Notice that,
for each d-d admissible colored partition
$\lam\in\Lambda_n^{(\gam)}$ with the fixed parity
$\rho(\lam)$,
the energy $\cE$, total momentum $\cP$ and
thermodynamic limit of the quantity $X_\lam Y_\lam^{-1}Z_\lam$
are invariant under the exchange $v_i\leftrightarrow v_j$
and/or $w_k\leftrightarrow w_l$.
Then, finally, after removing the order on momenta,
we arrive at the formula (\ref{dd-td}).
This formula coincides with our previous result
in \cite{KYA} up to the constant factor.

The formula (\ref{ss-td}) can be derived in the same way.

Essential part of
the formulae (\ref{rGreen-td}), (\ref{dd-td}) and (\ref{ss-td})
can be described by the function $F_\beta$.
As is the spinless case \cite{Ha},
we call the function $F_\beta$ the minimal form factor
of the $SU(2)$ spin CS model
(with integer coupling parameter).
The physical interpretation
of the minimal form factor has been discussed
in ref. \cite{KYA}.

\section{Conclusion}
\label{conclusion}

In this work,
we have taken the thermodynamic limit
of dynamical density and spin-density two-point
correlation functions of the spin CS model.
We have obtained the exact formulae
(\ref{dd-td}) and (\ref{ss-td})
of the density and spin-density two-point correlation
functions, respectively.
We have exactly shown that,
with appropriate numbers of quasiparticles and quasiholes,
the dynamical correlation functions of the spin CS model
can be described by
the unique function $F_\beta$ (\ref{form-factor})
which is called the minimal form factor.
These results are consistent
with our previous work \cite{KYA}.

\section*{Appendix A}
\label{appendixA}
In this appendix, we prove the following lemma:
for a colored partition
$\lam\in\Lambda_n$ with even (resp. odd) weight,
$S_\lam=0\quad(\mbox{resp. } \pm 1)
\Leftrightarrow
|W_\lam|=|H_{2}(\lam)|\quad
(\mbox{resp. } |H_{2}(\lam)|+1)$.
In this appendix, we do not assume that
$n$ is even.

We introduce the notations.
The partition $\lam=(\lam_1,\lam_2,\cdots\,)$
can be represented by the notation
$\lam=(1^{m_1(\lam)}2^{m_2(\lam)}\cdots\,)$ where
$m_i(\lam)=\#\{j\,|\,\lam_j=i\}$ \cite{Macd}.
Using this notation,
we define the following transformations
$\tau_i$ and $\tau'_i$ for $i\in\bZ_{>0}$:
\ba
&&
\tau_i:\,
\lam=(1^{m_1(\lam)}2^{m_2(\lam)}\cdots i^{m_i(\lam)}\cdots)
\longrightarrow
\left\{
\begin{array}{ll}
(1^{m_1(\lam)}2^{m_2(\lam)}\cdots i^{m_i(\lam)-2}\cdots),
\quad & m_i(\lam)\geq 2,
\\
\lam,
\quad & m_i(\lam)< 2,
\end{array}
\right.
\\
&&
\tau'_i:\,
\lam
\longrightarrow
(\tau_i(\lam'))'.
\ea
That is, $\tau_i$ ($\tau'_i$) is the following transformation:
if there exist two rows (columns)
which have same number of squares $i$
then $\tau_i$ ($\tau'_i$) removes these rows (columns),
if not then $\tau_i$ ($\tau'_i$) is the identity.
If $\lam$ has even (odd) weight
then both $\tau_i(\lam)$ and $\tau'_i(\lam)$
have even (odd) weights.
We introduce the special partition $\delta(k)\in\Lambda_n$ by
\be
\delta(k)
=
\left\{
\begin{array}{ll}
\O,
\quad & k=0,
\\
(k,k-1,\cdots,1)=(1^12^1\cdots k^1),
\quad & k=1,\cdots,n.
\end{array}
\right.
\ee
The partitions $\delta(k)$ for $k=0,1,\cdots,n$
are the fixed points of the transformations
$\tau_i$ and $\tau'_i$.
We see that, applying $\tau_i$'s and $\tau'_j$'s
sufficiently many times,
any partition $\lam\in\Lambda_n$ is mapped to
one of $\delta(k)$'s.
We denote the resultant mapping by
$\tau:\,\Lambda_n
\longrightarrow
\{\delta(k)\,|\,k=0,1,\cdots,n\}$.

{}From the definition,
the transformations $\tau_i$, $\tau'_i$ and $\tau$
can be defined on
the set of all colored partitions.
For instance,
$$
\tau
\left(
\,\,
\Yvcentermath1
\young(
\hfill\bullet\hfil\bullet,%
\bullet\hfil\bullet,%
\hfil\bullet\hfil,%
\bullet\hfil,%
\hfil\bullet,%
\bullet
)
\,\,
\right)
=
\Yvcentermath1
\young(
\hfill\bullet,%
\bullet
)
=\delta(2)
$$
We define the following numbers:
for a colored partition $\lam\in \Lambda_n$,
$wb(\lam)=|W_\lam|-|B_\lam|(=S_\lam)$ and
$wh(\lam)=|W_\lam|-|H_2(\lam)|$.
It is easy to see that
these numbers are invariant under the transformations
$\tau_i$ and $\tau'_i$,
{\it i.e.},
$wb(\tau_i(\lam))=wb(\lam)$,
$wb(\tau'_i(\lam))=wb(\lam)$ and same formulae
for $wh$.
Therefore
$wb(\tau(\lam))=wb(\lam)$ and $wh(\tau(\lam))=wh(\lam)$.
Moreover we have
\ba
&&
\label{wb}
wb(\delta(k))
=
\left\{
\begin{array}{rl}
0,
\quad & k=0,
\\
l,
\quad & k=2l-1,\,\quad(l=1,2,\cdots),
\\
-l,
\quad & k=2l,\,\quad(l=1,2,\cdots),
\end{array}
\right.
\\
&&
\label{wh}
wh(\delta(k))
=
\left\{
\begin{array}{rl}
0,
\quad & k=0,
\\
l^2,
\quad & k=2l-1,\,\quad(l=1,2,\cdots),
\\
l^2,
\quad & k=2l,\,\quad(l=1,2,\cdots).
\end{array}
\right.
\ea
(Notice that $H_2(\delta(k))=\O$ for all $k$).

We define the subset $\Lambda_n(\delta(k))$ of
$\Lambda_n$ by
\be
\Lambda_n(\delta(k))
=
\{
\lam\in\Lambda_n\,|\,
\tau(\lam)=\delta(k)
\}.
\ee
It is important to note the following fact:
if $\lam \in\Lambda_n(\delta(k))$ then
$wb(\lam)=wb(\delta(k))$ and $wh(\lam)=wh(\delta(k))$.
Therefore, from the formulae (\ref{wb}) and (\ref{wh}),
$\Lambda_n(\delta(k))\cap\Lambda_n(\delta(k'))=\O$
if $k\ne k'$.
This fact proves the lemma.

We have
the following decomposition
for the set of all {\it colored} partitions $\Lambda_n$
\be
\Lambda_n
=
\sqcup_{k=0}^n\Lambda_n(\delta(k)).
\ee
We see that
the set of all d-d (resp. s-s) admissible colored partitions is
the set $\Lambda_n(\delta(0))\cap\Lambda_n^{(\gam)}$
(resp.
$(\Lambda_n(\delta(1))\sqcup\Lambda_n(\delta(2)))
\cap\Lambda_n^{(\gam)}$).

\section*{Appendix B}
\label{appendixB}

In this appendix, we give examples of
the explicit formula for $Y_{\cA_\lam}$ and $Y_{\cB_\lam}$.

\bigskip
\noindent
a) Example for the d-d admissible colored partition

We consider the d-d admissible colored partition
$\lam\in\Lambda_n^{(\gam)}$ with parity
$\rho(\lam)
=(\overbrace{1,1,\cdots,1}^{\gam+1\ \mbox{{\footnotesize times}}})$.
In this case,
$I_1(\rho(\lam))=\{j\in I \,|\, j : \mbox{odd}\}$
and
$I_2(\rho(\lam))=\{j\in I \,|\, j : \mbox{even}\}$.

We have the explicit formula for $Y_{\cA_\lam}$ and $Y_{\cB_\lam}$
\ba
&&
Y_{\cA_\lam}
=
\xi^{-(\gam+1)}
\prod_{j\in I_1(\rho(\lam))}
(p/\gam+w_j(\lam')-1/2-\xi(j-2))
(p/\gam+w_j(\lam')-\xi(j-1)),
\\
&&
Y_{\cB_\lam}
=
\xi^{\gam+1}\Gam\big((1+1/\gam)/2\big)^{-\gam}
\prod_{j=1}^\gam\xi^{-2w_j(\lam')}
\nn\\
&&
\times
\prod_{j\in I_1(\rho(\lam))}
\Gam\big(w_j(\lam')-\xi(j-1)\big)\Gam\big(w_j(\lam')-\xi j+1/2\big)
\nn\\
&&
\times
\prod_{j\in I_2(\rho(\lam))}
\Gam\big(w_j(\lam')-\xi(j-1)+1/2\big)\Gam\big(w_j(\lam')-\xi j+1\big)
\nn\\
&&
\times
\prod_{{\scriptstyle j,k\in I_1(\rho(\lam))}
       \atop
       {\scriptstyle j<k}}
\frac{
\Gam\big(w_j(\lam')-w_k(\lam')+\xi(k-j)\big)
\Gam\big(w_j(\lam')-w_k(\lam')+\xi(k-j)-\xi+1/2\big)
}
{
\Gam\big(w_j(\lam')-w_k(\lam')+\xi(k-j)+\xi+1/2\big)
\Gam\big(w_j(\lam')-w_k(\lam')+\xi(k-j)+1\big)
}
\nn\\
&&
\times
\prod_{{\scriptstyle j,k\in I_2(\rho(\lam))}
       \atop
       {\scriptstyle j<k}}
\frac{
\Gam\big(w_j(\lam')-w_k(\lam')+\xi(k-j)\big)
\Gam\big(w_j(\lam')-w_k(\lam')+\xi(k-j)-\xi+1/2\big)
}
{
\Gam\big(w_j(\lam')-w_k(\lam')+\xi(k-j)+\xi+1/2\big)
\Gam\big(w_j(\lam')-w_k(\lam')+\xi(k-j)+1\big)
}
\nn\\
&&
\times
\prod_{{\scriptstyle j\in I_1(\rho(\lam)),k\in I_2(\rho(\lam))}
       \atop
       {\scriptstyle j<k}}
\frac{
\Gam\big(w_j(\lam')-w_k(\lam')+\xi(k-j)-1/2\big)
\Gam\big(w_j(\lam')-w_k(\lam')+\xi(k-j)-\xi\big)
}
{
\Gam\big(w_j(\lam')-w_k(\lam')+\xi(k-j)+\xi-1\big)
\Gam\big(w_j(\lam')-w_k(\lam')+\xi(k-j)-1/2\big)
}
\nn\\
&&
\times
\prod_{{\scriptstyle j\in I_2(\rho(\lam)),k\in I_1(\rho(\lam))}
       \atop
       {\scriptstyle j<k}}
\frac{
\Gam\big(w_j(\lam')-w_k(\lam')+\xi(k-j)+3/2\big)
\Gam\big(w_j(\lam')-w_k(\lam')+\xi(k-j)-\xi+2\big)
}
{
\Gam\big(w_j(\lam')-w_k(\lam')+\xi(k-j)+\xi+1\big)
\Gam\big(w_j(\lam')-w_k(\lam')+\xi(k-j)+3/2\big)
}.
\ea

\bigskip
\noindent
b) Example for the type I s-s admissible colored partition

We consider the type I s-s admissible colored partition
$\mu\in\Lambda_n^{(\gam)}$ with parity
$\rho(\mu)
=(\overbrace{1,1,\cdots,1}^{\gam\  \mbox{{\footnotesize times}}},0)$.
In this case,
$I_1(\rho(\mu))=\{j\in I \,|\, j : \mbox{even}\}$
and
$I_2(\rho(\mu))=\{j\in I \,|\, j : \mbox{odd}\}$.

The formula for
$Y_{\cA_\mu}$ is given by
\be
Y_{\cA_\mu}
=
\xi^{-(\gam-1)}
\prod_{j\in I_1(\rho(\mu))}
(p/\gam+w_j(\mu')+1/2-\xi(j-1))
(p/\gam+w_j(\mu')+1-\xi j).
\ee
The explicit form of $Y_{\cB_\lam}$ is given by the same formula in a)
with replacement of
$I_1(\rho(\lam))$ and $I_2(\rho(\lam))$
by $I_2(\rho(\mu))$ and $I_1(\rho(\mu))$, respectively.

\bigskip
\noindent
c) Example for the type II s-s admissible colored partition

Finally, we consider the type II
s-s admissible colored partition
$\eta\in\Lambda_n^{(\gam)}$ with parity
$$
\rho(\eta)
=
(\overbrace{0,1,0,1,\cdots,0,1,0}^{(\gam+3)/2}
\overbrace{0,1,0,1,\cdots,0,1}^{(\gam-3)/2},
0),
\quad
((\gam+3)/2:\, \mbox{odd}).
$$
In this case,
$I_1(\rho(\eta))=\{1,\cdots,\beta+2\}$
and
$I_2(\rho(\eta))=\{\beta+3,\cdots,2\beta+1\}$.

We have the explicit formulae
for $Y_{\cA_\eta}$ and $Y_{\cB_\eta}$
\ba
&&
Y_{\cA_\eta}
=
\xi^{-(\gam+3)}
\prod_{{\scriptstyle j\in I_1(\rho(\eta))}
       \atop
       {\scriptstyle j:\, \mbox{{\scriptsize odd}}}}
(p/\gam+w_j(\eta')-\xi(j-1))
(p/\gam+w_j(\eta')+1/2-\xi j)
\nn\\
&&\ \ \ \ \ \ \
\times
\prod_{{\scriptstyle j\in I_1(\rho(\eta))}
       \atop
       {\scriptstyle j:\, \mbox{{\scriptsize even}}}}
(p/\gam+w_j(\eta')+1/2-\xi(j-1))
(p/\gam+w_j(\eta')+1-\xi j),
\\
&&
Y_{\cB_\eta}
=
\xi^{\gam+5}\Gam\big((1+1/\gam)/2\big)^{-\gam}
\prod_{j=1}^\gam\xi^{-2w_j(\eta')}
\nn\\
&&
\times
\prod_{s=1,2}
\prod_{{\scriptstyle j\in I_s(\rho(\eta))}
       \atop
       {\scriptstyle j:\, \mbox{{\scriptsize odd}}}}
\Gam\big(w_j(\eta')-\xi(j-1)\big)\Gam\big(w_j(\eta')-\xi j+1/2\big)
\nn\\
&&
\times
\prod_{s=1,2}
\prod_{{\scriptstyle j\in I_s(\rho(\eta))}
       \atop
       {\scriptstyle j:\, \mbox{{\scriptsize even}}}}
\Gam\big(w_j(\eta')-\xi(j-1)+1/2\big)\Gam\big(w_j(\eta')-\xi j+1\big)
\nn\\
&&
\times
\prod_{s=1,2}
\prod_{{{\scriptstyle j,k\in I_s(\rho(\eta))}
         \atop
        {\scriptstyle j<k}}
        \atop
       {\scriptstyle j,k:\, \mbox{{\scriptsize odd}}}}
\frac{
\Gam\big(w_j(\eta')-w_k(\eta')+\xi(k-j)\big)
\Gam\big(w_j(\eta')-w_k(\eta')+\xi(k-j)-\xi+1/2\big)
}
{
\Gam\big(w_j(\eta')-w_k(\eta')+\xi(k-j)+\xi+1/2\big)
\Gam\big(w_j(\eta')-w_k(\eta')+\xi(k-j)+1\big)
}
\nn\\
&&
\times
\prod_{s=1,2}
\prod_{{{\scriptstyle j,k\in I_s(\rho(\eta))}
         \atop
        {\scriptstyle j<k}}
        \atop
       {\scriptstyle j,k:\, \mbox{{\scriptsize even}}}}
\frac{
\Gam\big(w_j(\eta')-w_k(\eta')+\xi(k-j)\big)
\Gam\big(w_j(\eta')-w_k(\eta')+\xi(k-j)-\xi+1/2\big)
}
{
\Gam\big(w_j(\eta')-w_k(\eta')+\xi(k-j)+\xi+1/2\big)
\Gam\big(w_j(\eta')-w_k(\eta')+\xi(k-j)+1\big)
}
\nn\\
&&
\times
\prod_{s=1,2}
\prod_{{{\scriptstyle j,k\in I_s(\rho(\eta))}
         \atop
        {\scriptstyle j<k}}
        \atop
        {\scriptstyle j:\, \mbox{{\scriptsize odd}},\
                     k:\, \mbox{{\scriptsize even}}}}
\frac{
\Gam\big(w_j(\eta')-w_k(\eta')+\xi(k-j)-1/2\big)
\Gam\big(w_j(\eta')-w_k(\eta')+\xi(k-j)-\xi\big)
}
{
\Gam\big(w_j(\eta')-w_k(\eta')+\xi(k-j)+\xi\big)
\Gam\big(w_j(\eta')-w_k(\eta')+\xi(k-j)+1/2\big)
}
\nn\\
&&
\times
\prod_{s=1,2}
\prod_{{{\scriptstyle j,k\in I_s(\rho(\eta))}
        \atop
        {\scriptstyle j<k}}
        \atop
        {\scriptstyle j:\, \mbox{{\scriptsize odd}},\
                     k:\, \mbox{{\scriptsize even}}}}
\frac{
\Gam\big(w_j(\eta')-w_k(\eta')+\xi(k-j)+1/2\big)
\Gam\big(w_j(\eta')-w_k(\eta')+\xi(k-j)-\xi+1\big)
}
{
\Gam\big(w_j(\eta')-w_k(\eta')+\xi(k-j)+\xi+1\big)
\Gam\big(w_j(\eta')-w_k(\eta')+\xi(k-j)+3/2\big)
}
\nn\\
&&
\times
\prod_{{\scriptstyle j\in I_1(\rho(\eta)),
                      k\in I_2(\rho(\eta))}
        \atop
        {\scriptstyle j,k:\, \mbox{{\scriptsize odd}}}}
\frac{
\Gam\big(w_j(\eta')-w_k(\eta')+\xi(k-j)+1\big)
\Gam\big(w_j(\eta')-w_k(\eta')+\xi(k-j)-\xi+3/2\big)
}
{
\Gam\big(w_j(\eta')-w_k(\eta')+\xi(k-j)+\xi+1/2\big)
\Gam\big(w_j(\eta')-w_k(\eta')+\xi(k-j)+1\big)
}
\nn\\
&&
\times
\prod_{{\scriptstyle j\in I_1(\rho(\eta)),
                      k\in I_2(\rho(\eta))}
        \atop
        {\scriptstyle j,k:\, \mbox{{\scriptsize even}}}}
\frac{
\Gam\big(w_j(\eta')-w_k(\eta')+\xi(k-j)+1\big)
\Gam\big(w_j(\eta')-w_k(\eta')+\xi(k-j)-\xi+3/2\big)
}
{
\Gam\big(w_j(\eta')-w_k(\eta')+\xi(k-j)+\xi+1/2\big)
\Gam\big(w_j(\eta')-w_k(\eta')+\xi(k-j)+1\big)
}
\nn\\
&&
\times
\prod_{{\scriptstyle j\in I_1(\rho(\eta)),
                      k\in I_2(\rho(\eta))}
        \atop
        {\scriptstyle j:\, \mbox{{\scriptsize odd}},\,
                      k:\, \mbox{{\scriptsize even}}}}
\frac{
\Gam\big(w_j(\eta')-w_k(\eta')+\xi(k-j)+1/2\big)
\Gam\big(w_j(\eta')-w_k(\eta')+\xi(k-j)-\xi+1\big)
}
{
\Gam\big(w_j(\eta')-w_k(\eta')+\xi(k-j)+\xi\big)
\Gam\big(w_j(\eta')-w_k(\eta')+\xi(k-j)+1/2\big)
}
\nn\\
&&
\times
\prod_{{\scriptstyle j\in I_1(\rho(\eta)),
                      k\in I_2(\rho(\eta))}
        \atop
       {\scriptstyle j:\, \mbox{{\scriptsize even}},\,
                      k:\, \mbox{{\scriptsize odd}}}}
\frac{
\Gam\big(w_j(\eta')-w_k(\eta')+\xi(k-j)+3/2\big)
\Gam\big(w_j(\eta')-w_k(\eta')+\xi(k-j)-\xi+2\big)
}
{
\Gam\big(w_j(\eta')-w_k(\eta')+\xi(k-j)+\xi+1\big)
\Gam\big(w_j(\eta')-w_k(\eta')+\xi(k-j)+3/2\big)
}.
\ea

\noindent {\bf Acknowledgement:}

The authors thank Y. Kato and Y. Kuramoto
for discussions.
TY is grateful to K. Takemura for useful comments.
TY was supported by
the Core Research for Evolutional Science and Technology (CREST)
program of the Science and Technology Agency of Japan.

\vspace{24pt}



\end{document}